\documentclass[conference]{IEEEtran}
\IEEEoverridecommandlockouts

\RequirePackage[absolute]{textpos}
\DeclareRobustCommand*{\copyrightnote}{%
  \begin{textblock}{85}(17.5,256.75)
      \scriptsize{\noindent \copyright 2021 IEEE. Personal use of this material is permitted. Permission from IEEE must be obtained for all other uses, in any current or future media, including reprinting/republishing this material for advertising or promotional purposes, creating new collective works, for resale or redistribution to servers or lists, or reuse of any copyrighted
component of this work in other works.}%
  \end{textblock}%
    }
    
\usepackage[
    colorlinks=true,
    linkcolor={blue!65!black},
    citecolor={blue!65!black},
    filecolor=black,
    urlcolor={blue!65!black},
    plainpages=false,
    pdfpagelabels,
    breaklinks=true
    ]{hyperref}
    
\usepackage{amsmath,amssymb,amsfonts}
\usepackage{algorithmic}
\usepackage{graphicx}
\usepackage{textcomp}
\usepackage[dvipsnames]{xcolor}
\definecolor{vsusp}{HTML}{D62828}
\definecolor{susp}{HTML}{FF931F}
\definecolor{unexp}{HTML}{6D3764}
\definecolor{exp}{HTML}{007CBE}
\usepackage{bm}
\usepackage{caption}
\usepackage{subcaption}
\usepackage{multirow}
\usepackage{multicol}
\usepackage[nottoc]{tocbibind}
\usepackage{tabularx}

\usepackage[giveninits=true,sorting=none]{biblatex}
\AtEveryBibitem{\clearfield{doi}}
\AtEveryBibitem{\clearfield{pages}}

\usepackage[hang,flushmargin]{footmisc}

\def\BibTeX{{\rm B\kern-.05em{\sc i\kern-.025em b}\kern-.08em
    T\kern-.1667em\lower.7ex\hbox{E}\kern-.125emX}}
    
\IEEEoverridecommandlockouts
\IEEEpubid{\makebox[\columnwidth]{978-1-7281-5684-2/20/\$31.00~\copyright{}2021 IEEE \hfill} \hspace{\columnsep}\makebox[\columnwidth]{ }}

\begin{document}

\title{Machine learning on knowledge graphs for context-aware security monitoring\\
\thanks{
This work was partially funded by the Federal Ministry for Economic Affairs and Energy of Germany (BMWi) within the IIP-Ecosphere Project.
}
\thanks{
\dag These authors contributed equally to this work.}
}
\author{\IEEEauthorblockN{Josep Soler Garrido$^\dag$, Dominik Dold$^\dag$ and Johannes Frank}
\IEEEauthorblockA{\textit{Siemens Technology} \\
Munich, Germany \\
\{josep.soler\_garrido, dominik.dold, johannes.frank\}@siemens.com}
}

\maketitle
\copyrightnote
\thispagestyle{plain}
\pagestyle{plain}

\begin{abstract}
Machine learning techniques are gaining attention in the context of intrusion detection due to the increasing amounts of data generated by monitoring tools, as well as the sophistication displayed by attackers in hiding their activity. However, existing methods often exhibit important limitations in terms of the quantity and relevance of the generated alerts. Recently, knowledge graphs are finding application in the cybersecurity domain, showing the potential to alleviate some of these drawbacks thanks to their ability to seamlessly integrate data from multiple domains using human-understandable vocabularies. We discuss the application of machine learning on knowledge graphs for intrusion detection and experimentally evaluate a link-prediction method for scoring anomalous activity in industrial systems. After initial unsupervised training, the proposed method is shown to produce intuitively well-calibrated and interpretable alerts in a diverse range of scenarios, hinting at the potential benefits of relational machine learning on knowledge graphs for intrusion detection purposes.
\end{abstract}


\section{Introduction}
Artificial Intelligence (AI) methods and techniques such as Machine Learning (ML) are bound to play an increasingly prominent role in the cybersecurity domain \cite{nstc}. Potential applications are found both in defensive and offensive scenarios across most cybersecurity functions \cite{Truong,2018framework}. Among these, detection tasks are particularly poised to benefit from the ability to automatically analyze and learn from vast quantities of data. Indeed, there are many relevant examples of the application of deep learning and similar techniques for intrusion detection systems (IDS) \cite{ALDWEESH2020105124}, based either on supervised learning techniques targeting detection of known malicious signatures, or on anomaly detection methods able to detect unknown attack patterns and find deviations from a previously learned baseline.  
However, it has been argued that, while advantageous in terms of detection capabilities and data requirements, the latter class of machine learning-based IDS systems must address some drawbacks in order to improve their usefulness \cite{ALDWEESH2020105124}, including, among others, their tendency to produce an unmanageable amount of alerts \cite{Truong, 8909930}, their limited ability to learn from and interact with human operators \cite{8909930, 7502263}, and the lack of interpretability of the alerts generated \cite{9069273}. 

Addressing these drawbacks would be of great benefit not only to defend conventional IT systems, but also in the context of modern operational technology (OT) systems, such as those used in factories and other industrial automation settings. Indeed, while industrial control systems have traditionally been extremely deterministic in their operation---and their security has been at least partially implemented by means of isolation \cite{BHAMARE2020101677}---modern Industry 4.0 automation settings are characterized by a convergence of IT and OT infrastructure \cite{atos}, displaying increasingly complex activity patterns and network topologies and making extensive use of autonomous systems and components (e.g. robots, or AI-enabled software applications) that constantly interact with one another. This new paradigm has the potential to substantially improve the flexibility, reliability and efficiency of industrial systems, but also poses new cybersecurity challenges \cite{8938806} and demands a high degree of domain-specific knowledge from analysts assessing potential integrity issues or indications of security compromises.  

It is in this context that we propose the application of machine learning on knowledge graphs in order to improve the quality and relevance of IDS-generated alerts in modern industrial systems, increasing their usefulness for human operators. Intuitively, this can be achieved by leveraging the semantic integration of domain-specific data on a single knowledge base, which enables us to better contextualize and enrich cybersecurity-relevant observations, and allows machine learning methods to leverage this additional context, i.e., to learn from these observations in a way that makes use of the rich set of interconnections and relations between different entities in the graph. In our case, this is shown to result in a model of the behavior of the different actors in an industrial automation system that exhibits particularly useful generalization properties and can be employed to reliably detect and score unexpected activity patterns. 

The rest of the paper is organized as follows. Section \ref{sec:kg} introduces knowledge graphs and presents a brief overview of existing applications in the cybersecurity domain. In section \ref{sec:application} we describe the graph learning algorithm employed for our specific security monitoring application. Section \ref{sec:setup} describes an empirical evaluation setup based on an industrial automation hardware prototype. The experiments performed and results obtained are presented and discussed in section \ref{sec:results}, before finally drawing some conclusions in section \ref{sec:conclusions}.

\section{Background and related work}
\label{sec:kg}
\subsection{Knowledge graph model}
\begin{figure}[t]
    \centering
    \includegraphics[width=\columnwidth]{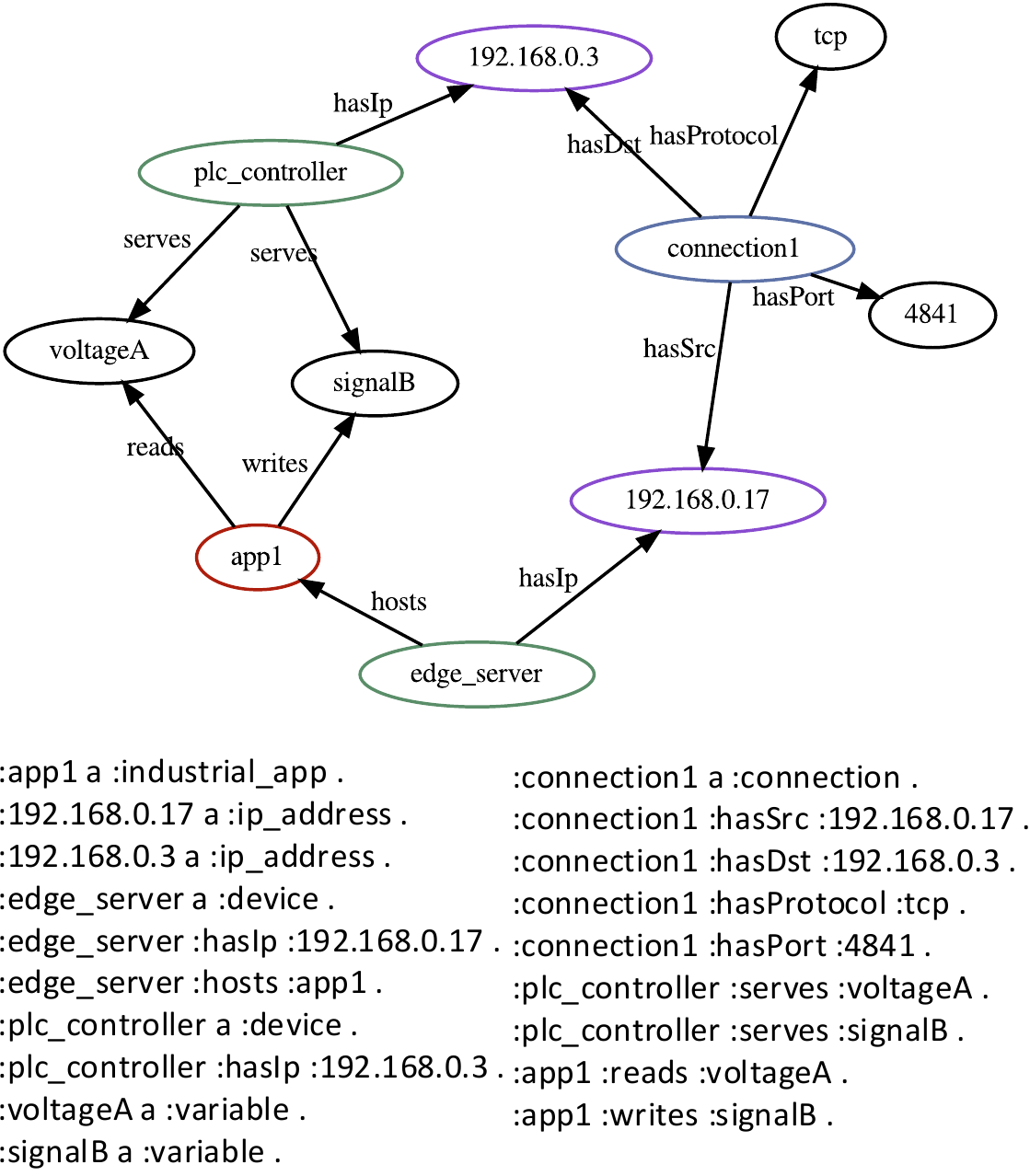}
     \caption{Simple graph describing some elements in an industrial network, and list of statements represented in the graph.}
	\label{fig1}
\end{figure}

A knowledge graph is a specific type of knowledge base where information is encoded in the form of a directed labeled graph, with nodes representing entities and edges representing different types of possible relationships between entities. The information contained in a knowledge graph can be alternatively represented as a list of statements with the form subject-predicate-object, or $\{s,p,o\}$ for short, where $s$ corresponds to an entity in the graph, $p$ to a specific relationship, and $o$ can either be an entity or a literal, e.g., a specific data value. A simple knowledge graph illustrating these two alternative representations is depicted in Figure \ref{fig1}. 

Knowledge graphs are particularly useful structures to integrate data from multiple areas of knowledge, typically making use of domain-specific vocabularies and ontologies that model the different categories, relationships, rules and constraints in a specific area of knowledge. Sophisticated data discovery and analytics tasks can be performed on the resulting graphs, e.g., by means of semantic queries using languages such as SPARQL in RDF graphs \cite{sparql}. Use of machine learning methods is also possible on knowledge graphs, typically by means of so-called node embeddings: vector representations of graph entities which are more suitable for processing via neural networks and similar methods than their original symbolic representations. 

Many different methods have been proposed to learn embeddings \cite{8294302}. A specific method used in the context of our work is RESCAL \cite{rescal}. Assuming a knowledge graph with $n$ entities and $m$ different types of relations, a suitable representation is given by $\bf{X}$, a binary tensor of size $n \times m \times n$ where entry $X_{s,p,o}$ contains a $1$ if the corresponding triple $\{s,p,o\}$ is present in the graph and $0$ otherwise. Given such a representation, RESCAL formulates the graph embedding problem as a tensor factorization operation
\begin{equation}
\label{eq:rescal}
    X_{s,p,o}\,{\approx}\,\bm{e}^T_s\bm{R}_p\bm{e}_o ,
\end{equation}
where $\bm{e}_j$ is the $R \times 1$ embedding vector of entity $j$ and $\bm{R}_k$ is a $R \times R$ matrix modelling the interactions between entity embeddings for the k-th relationship type. 

Embeddings are obtained through minimization of the reconstruction loss
\begin{equation}
    L_\mathrm{MSE} = \sum_{s,p,o} \|X_{s,p,o} - \bm{e}^T_s\bm{R}_p\bm{e}_o \|^2 \,,
\end{equation}
which can be performed, for instance, by stochastic gradient descent (SGD), sequentially sampling batches of positive and negative (i.e., not present in the graph) triple statements in order to iteratively update the learned representations. Once the learning step is completed, the resulting embeddings can be used for a wide range of downstream applications. For instance, in a knowledge graph completion task, non-observed relationships can be inferred. In the RESCAL case this is done by means of the score $\theta_{s,p,o} = \bm{e}^T_s\bm{R}_p\bm{e}_o$, which captures the likelihood that, given the set of observed triples, $\{s,p,o\}$ also represents a true fact.

\begin{figure*}[!ht]
\centering
    \begin{subfigure}[t]{0.30\textwidth}
        \includegraphics[height=2.65in]{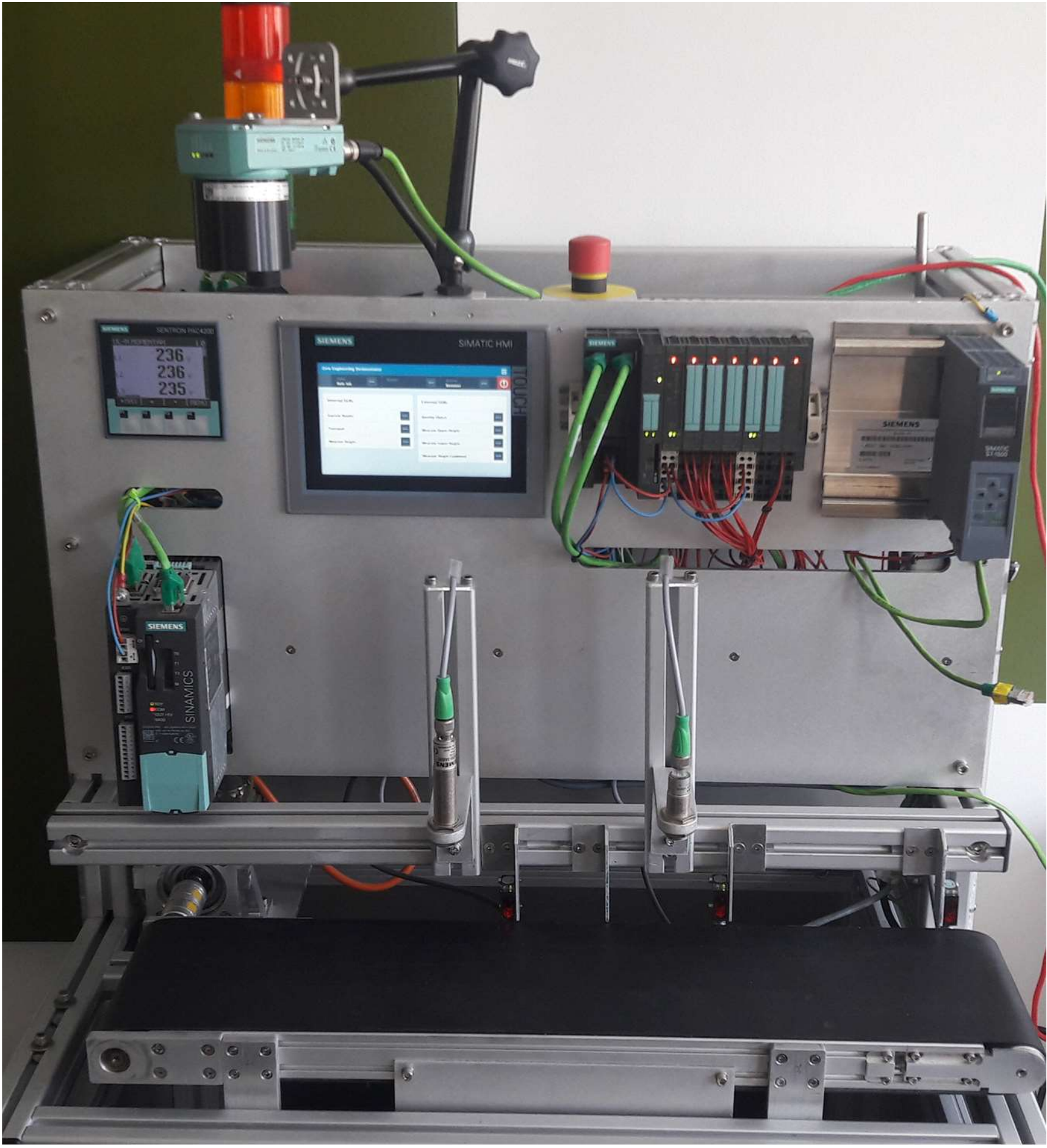}
    \end{subfigure}%
    \hfill \hfill
    \begin{subfigure}[t]{0.65\textwidth}
        \includegraphics[height=2.65in]{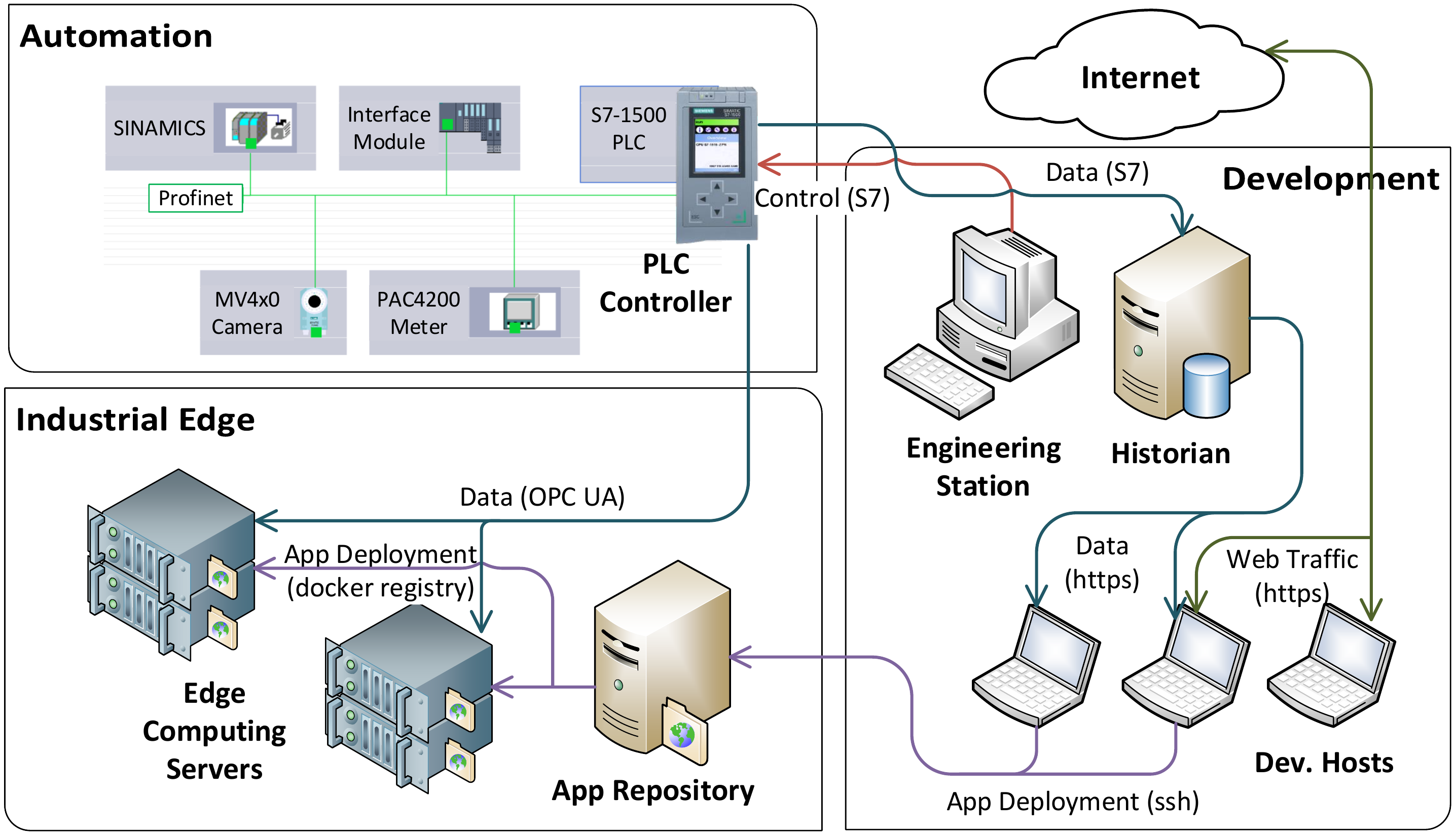}
    \end{subfigure}
    \caption{Hardware demonstrator used for evaluation (left) and simplified network diagram including main traffic flows (right).  }
    \label{fig:prototype}
\end{figure*}

\subsection{Cybersecurity applications}
\label{sec:cybersecurity}
Analysis of security incidents typically requires consideration of multiple data sources, some of which are often exchanged between organizations. In order to facilitate this, common schemas and data representation formats, such as STIX \cite{stix}, have been introduced that enable organizations to exchange threat intelligence in a consistent way. More recently, these have evolved into fully-fledged ontologies enabling inference and reasoning \cite{Syed2016UCOAU}. These ontologies model a wide range of cybersecurity-relevant knowledge such as product information, known vulnerabilities and attack patterns, and can additionally be linked to domain-specific knowledge, e.g. coming from industrial automation systems \cite{9237126}.

Construction of high-quality knowledge graphs is a challenging task, especially when it requires extraction of information from unstructured textual or heterogeneous data. Therefore, the use of machine learning to perform automatic extraction of cybersecurity knowledge from various sources has been an area of strong academic interest \cite{8537852, 9073093}. 

Once constructed, these knowledge graphs find a wide range of applications, e.g., vulnerability assessment \cite{sepses}, intrusion and threat detection \cite{sepses,8537852,malrank}, malware analysis \cite{9264152}, and other forms of AI-enabled assistance to security analysts \cite{cyberallintel}. 
These applications make use of various methods in order to extract insights from cybersecurity knowledge graphs, ranging from simple SPARQL queries to complex rule-based systems, ranking algorithms or logical reasoning methods. 

\section{Learning method}
\label{sec:application}
We apply machine learning on knowledge graphs to detect unexpected activity in industrial automation systems. Multiple studies have considered detection of anomalies using graph data \cite{a19019efc43b4c0cab1fb8c22b8e040e}. More recently, several of these methods have focused on multi-relational data, as found in knowledge graphs \cite{ijcai2019-614,9020760,8455737}. In fact, knowledge graph completion methods, able by definition to infer missing facts, can be readily applied to anomaly detection problems \cite{kgrefinementpaulheim}. That is the approach we follow, learning a baseline of normal behavior by training a generative graph embedding algorithm on a graph built from a training dataset, and using it thereafter in a link-prediction setting to rank the likelihood of triple statements resulting from events observed at test time and determine whether they represent a significant deviation from the baseline.

We employ a tensor factorization method inspired by RESCAL. In our case, a novel probabilistic formulation is employed \cite{dold2}, where the joint probability of a graph represented by $\bm{X}$ is modelled as
\begin{align}
    p(\pmb{X}) = \frac{1}{Z}\, e^{- E(\pmb{X})} \,,
\end{align}
where $E(\pmb{X})$ is an energy function given by
\begin{equation}
    E(\pmb{X}) = -\sum_{s,p,o} X_{s,p,o}\, \theta_{s,p,o} \,,
\end{equation}
and $Z$ is a normalization constant given by a summation over all possible graph realizations
\begin{equation}
    Z = \sum_{\pmb{X}'}  e^{- E(\pmb{X}')} \,.
\end{equation}
Such a stochastic description of graph data allows us to handle data streams where identical events appear multiple times or underlie statistical variations---something that is not covered by traditional graph embedding algorithms like RESCAL.
By employing wake-sleep learning \cite{hinton1995wake}, where the model learns in an unsupervised mode by only observing positive triples, embeddings are found that assign low energies to triple statements consistent with the training graph, and high energies otherwise.
This energy-based formulation is experimentally shown to produce a broad range of triple scores $\theta_{s,p,o}$ compared to standard RESCAL factorization and other similar graph embedding methods, which in our case have been observed to result in a more bimodal distribution of scores, i.e., they tend to assign either very low or very high scores to potentially anomalous triples. Having a system that produces a wide range of scores is beneficial in the proposed IDS context, as it allows human analysts to better prioritize alerts. These benefits are further illustrated in the next sections which describe a thorough experimental evaluation campaign. 

\section{Experimental setup}
\label{sec:setup}
The hardware prototype is described in Figure \ref{fig:prototype}, following the design of modern industrial systems integrating IT and OT elements. The automation side is equipped with a Siemens S7-1500 PLC connected to peripherals via an industrial network. These include a drive subsystem controlling the motion of a conveyor belt, an industrial camera, a human-machine-interface (HMI), and a distributed I/O subsystem with modules interfacing with various sensors for object positioning and other measurements. The PLC exposes values reported by these sensors as well as information about the state of the system by means of an OPC-UA server \cite{mahnke2009opc}. 

The variables exposed by the server are consumed on the IT part of the demonstrator by applications hosted on edge computing servers, i.e., computing infrastructure directly located at the shop floor which is typically devoted to data-driven tasks that require close integration with the automation system and short response times, such as real-time system monitoring, fault prediction or production optimization. 

Industrial edge applications have dynamic lifecycles, and this is captured in the prototype by recreating a development environment. This cycle starts with development hosts consuming potentially high volumes of data from a historian, a database that constantly stores process data from the automation system. Thereafter, containerized applications---ostensibly developed on the basis of the consumed data---are pushed into an application repository. Finally, edge computing hosts fetch application updates periodically. To make the behavior more realistic, development hosts occasionally access the internet with low traffic volumes. The environment is fully virtualized and performs these activities in an autonomous manner, with an option to manually induce different types of anomalous behaviors in order to test the response of our IDS system.

A knowledge graph is built out of the running prototype by integrating three main sources of knowledge: information about the automation system, observations at the network level (e.g., connections between hosts), and observations at the application level (e.g., data access events). A sizeable portion of the information is related to the automation system, which is extracted from engineering tools in the AutomationML format and ingested into the graph using a readily available ontology \cite{7301643}. Information about application activity is obtained from the OPC-UA server logs, including session information, which variables are accessed and in which way. Finally, all network traffic is passed through the security monitoring tool Zeek \cite{zeek}, which produces a stream of observed connections that are ingested using a simple custom data model. 

\newcommand{\scoreObserved}{Observed}
\newcommand{\scoreTwo}{{\color{exp}Expected}}
\newcommand{\scoreOneTwo}{{\color{unexp}Unexpected}}
\newcommand{\scoreOne}{{\color{susp}Suspicious}}
\newcommand{\scoreZero}{{\color{vsusp}Highly Suspicious}}

\begin{figure}[!h]
\centering
    \begin{subfigure}[t]{0.5\textwidth}
        \centering
        \includegraphics[height=2.5in]{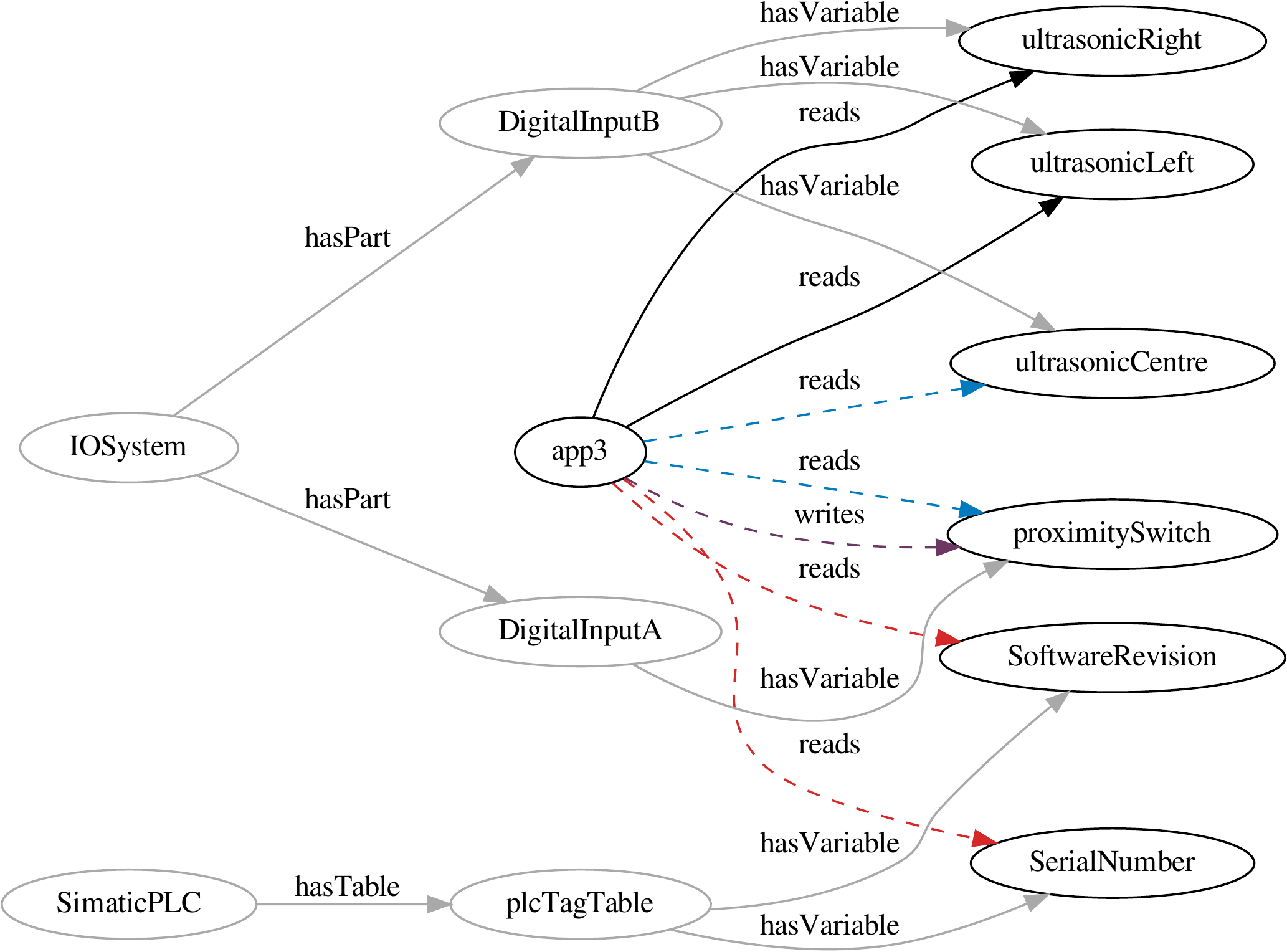}
        \caption{sub-graph containing variable access events}
    \end{subfigure}%
    \vspace{0.1in}
    \begin{subfigure}[b]{0.5\textwidth}
        \centering
        \includegraphics[height=3.7in]{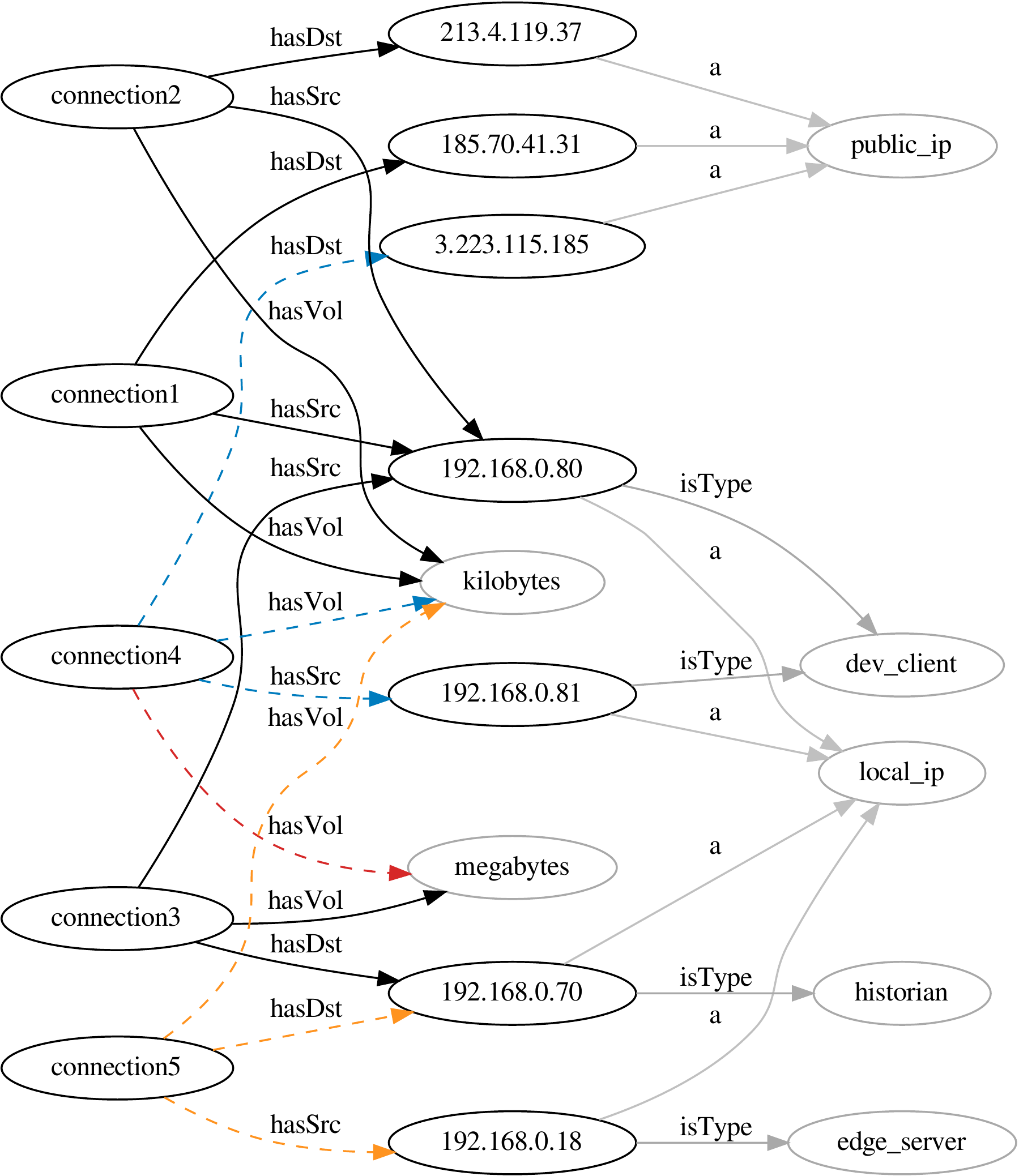}
        \caption{sub-graph containing network connections}
    \end{subfigure}
    \caption{Illustration of two evaluation scenarios, including baseline activity observed during training (black edges) and deviations of different severity (dashed colored edges: \scoreTwo, \scoreOneTwo, \scoreOne, \scoreZero ). {\bf (a)} Expected scores for data-access events, depending on the context provided by observed ones, e.g., which modules or subsystems previously accessed variables are related to. {\bf (b)} Expected scores for new HTTPS connections (4--5), depending on the context provided by observed ones (1--3), e.g. whether hosts of the same type are involved in similar traffic patterns.} 
\label{fig:graph_examples}
\end{figure}

\begin{table*}[ht]
\centering
        \begin{tabular}{|l|ll|}
            \multicolumn{1}{l}{\bf{Test case}} & \multicolumn{1}{l}{\bf{Scenario description}} & \multicolumn{1}{l}{\bf{Expected severity}}\\
            \hline
            \multirow{4}{*}{Variable access} & \multicolumn{1}{l}{App accesses the same variables as during training.} & \multicolumn{1}{l|}{\scoreObserved} \\\cline{2-3}
                                 & \multicolumn{1}{l}{App accesses variables from the same (or a closely related) device or module as those observed during training.} & \multicolumn{1}{l|}{\scoreTwo}  \\\cline{2-3}
                                 & \multicolumn{1}{l}{App changes the way it accesses some variables (e.g. writes instead of reads).} & \multicolumn{1}{l|}{\scoreOneTwo} \\\cline{2-3}
                                 & \multicolumn{1}{l}{App accesses variables completely unrelated to those accessed during training.} & \multicolumn{1}{l|}{\scoreZero} \\
            \hline \hline
            \multirow{6}{*}{HTTPS access} & \multicolumn{1}{l}{ A dev. host makes an HTTPS access (internet or historian) as observed during training for the same host.} & \multicolumn{1}{l|}{\scoreObserved} \\\cline{2-3}
                                 & \multicolumn{1}{l}{A dev. host makes an HTTPS access for the first time, but in a way consistent with the actions of other dev. hosts.} & \multicolumn{1}{l|}{\scoreTwo}  \\\cline{2-3}
                                 & \multicolumn{1}{l}{A dev. host observed to access the internet during training accesses a previously unseen public IP address.} & \multicolumn{1}{l|}{\scoreTwo} \\\cline{2-3}
                                 & \multicolumn{1}{l}{A local address not corresponding to a dev. host (e.g. an edge server) accesses the historian.} & \multicolumn{1}{l|}{\scoreOne} \\\cline{2-3}
                                 & \multicolumn{1}{l}{A local address not corresponding to a dev. host (e.g. an edge server) accesses a public IP address.} & \multicolumn{1}{l|}{\scoreZero} \\\cline{2-3}
                                 & \multicolumn{1}{l}{A high-volume HTTP access is made to a public IP address (high volumes only from historian in baseline).} & \multicolumn{1}{l|}{\scoreZero} \\
            \hline \hline
            \multirow{6}{*}{SSH access} & \multicolumn{1}{l}{ A dev. host makes an SSH access to the app repository as seen during training.} & \multicolumn{1}{l|}{\scoreObserved} \\\cline{2-3}
                                 & \multicolumn{1}{l}{A dev. host makes an SSH access to the app repository for the first time, but exactly like other dev. hosts.} & \multicolumn{1}{l|}{\scoreTwo}  \\\cline{2-3}
                                 & \multicolumn{1}{l}{A dev. host accesses the app repository via SSH as during training but with a slightly higher data volume.} & \multicolumn{1}{l|}{\scoreTwo} \\\cline{2-3}
                                 & \multicolumn{1}{l}{The historian host (not a dev. host but on the same network) accesses the app repository via SSH.} & \multicolumn{1}{l|}{\scoreOneTwo} \\\cline{2-3}
                                 & \multicolumn{1}{l}{A dev. host accesses an edge server via SSH, but during training no edge servers received SSH connections.} & \multicolumn{1}{l|}{\scoreOne} \\\cline{2-3}
                                 & \multicolumn{1}{l}{SSH connection between two edge servers. During training no edge servers started or received SSH connections.} & \multicolumn{1}{l|}{\scoreZero} \\
            \hline \hline
            \multirow{3}{*}{Credential use} & \multicolumn{1}{l}{Access to OPC-UA server from the same IP address as observed for the corresponding app during training.} & \multicolumn{1}{l|}{\scoreObserved} \\\cline{2-3}
                                 & \multicolumn{1}{l}{Access to OPC-UA server from a different IP address but which also corresponds to an edge server.} & \multicolumn{1}{l|}{\scoreTwo}  \\\cline{2-3}
                                 & \multicolumn{1}{l}{Access to OPC-UA server from an IP address that corresponds to a development host.} & \multicolumn{1}{l|}{\scoreOne} \\
            \hline \hline
            \multirow{4}{*}{Network scan} & \multicolumn{1}{l}{Connections (source, dest., port) matching those observed during training. } & \multicolumn{1}{l|}{\scoreObserved} \\\cline{2-3}
                                 & \multicolumn{1}{l}{Connection matching source-destination pairs observed during training, but on a different port.} & \multicolumn{1}{l|}{\scoreTwo}  \\\cline{2-3}
                                 & \multicolumn{1}{l}{Connection which does not match any source-destination pair observed during training.} & \multicolumn{1}{l|}{\scoreOne} \\\cline{2-3}
                                 & \multicolumn{1}{l}{Attempt to connect to an IP which is not assigned to any host.} & \multicolumn{1}{l|}{\scoreZero} \\
            \hline
        \end{tabular}
    \caption{List of scenarios and baseline deviations generated for evaluation, and their a-priori severity assigned.}
    \label{table:scenarios}
\end{table*}

\begin{figure*}[t]
    \centering
    \includegraphics[width=0.95\textwidth]{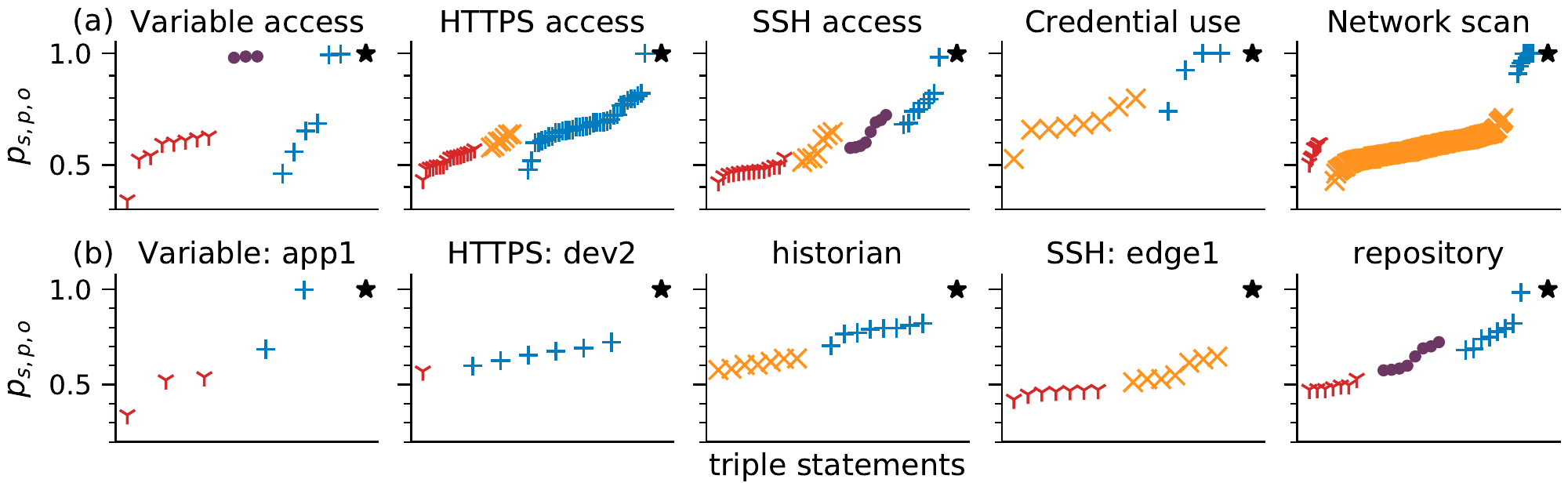}
	\caption{Probabilities---derived from scores---for test-set triples corresponding to (a) the different scenarios described in Table \ref{table:scenarios} and (b) specific graph entities, from left to right: variables accessed by application \#1, HTTPS events performed by development host \#2, HTTPS events with the historian as destination, SSH events involving edge computing host \#1 and SSH events involving the app repository. Markers and colors denote the expected severity: \scoreZero\ {\color{vsusp}(Y)}, \scoreOne\ {\color{susp}($\times$)}, \scoreOneTwo\ {\color{unexp}($\bullet$)}, \scoreTwo\ {\color{exp}($+$)} and Observed ($\star$). To increase readability, only the average probability of Observed triples is shown.} 
	\label{fig:experiments}
\end{figure*}

\section{Evaluation}
\label{sec:results}
\subsection{Methodology}

Initially, a baseline is captured with the system under normal operating conditions, and the collected data is used to train the link prediction algorithm of section \ref{sec:application} in an unsupervised manner. Thereafter, in order to qualitatively evaluate its predictions, we trigger a set of actions which result in events not observed during normal operation, but which would be assigned a wide range of severity levels by a human expert upon detailed analysis of the available contextual information.

Table \ref{table:scenarios} contains the full list of scenarios evaluated, ranging from subtle baseline deviations which are likely benign in nature to more severe ones which may be indicative of malicious activity such as reconnaissance, app and host compromise, credential stealing, lateral movement or data exfiltration. 

Some of these scenarios, related to app and network activity, are illustrated in Figure \ref{fig:graph_examples}. The examples highlight different expected scores for specific statements based on observed activity, e.g. whether similar apps or hosts behave in similar ways. It should be noted that any single event, e.g. a connection between two hosts, can result in multiple statements on the graph, with potentially different associated scores. For example, source, destination and port in a connection may be ranked as expected while the edges corresponding to traffic volume or to the service name may be flagged as suspicious, giving the resulting alerts a high degree of interpretability when examined by a human security analyst.

\subsection{Results}

Results for the different test cases are shown in Figure \ref{fig:experiments}a in probabilistic form by applying the logistic function to obtained scores $p_{s,p,o} = \sigma\left(\theta_{s,p,o}\right)$. In general, it is observed that they follow the expected trend, with empirical score distributions showing lower expected values for scenarios with higher severity levels. Perhaps unsurprisingly, there are some exceptions, e.g., the statements related to variable write access which are ranked higher than initially expected. However, this can be due to the fact that write operations are rather infrequent in our dataset, making it more difficult for the algorithm to learn the general expected behaviour. 

It should be noted that data modelling choices, i.e., how the raw captured data is represented in the form of entities and relationships, can have an impact on the resulting scores. This means that some representation choices can be beneficial for certain scenarios and detrimental to others. Despite this, for practical reasons, our evaluation campaign was carried out using a single model trained on a single graph and tested on all the different scenarios.

Results for statements involving specific graph entities are shown in Figure \ref{fig:experiments}b. Compared to the previous aggregated plots, these typically show a clearer trend, i.e., a lower degree of score overlap for different severity levels. This indicates that it may be beneficial for analysts using our IDS system to compare the observed scores with individual expected baselines for the specific entities involved, rather than defining general thresholds for all events to assess their severity.

Finally, it should be noted that all the events generated by the system prototype at test-time are represented in the figures for the corresponding test cases. This includes any statements resulting from spurious events unseen during training but unrelated to the test-cases themselves. These can result from non-deterministic behaviors built into the different software components or unexpected network issues, and typically represent small deviations from the baseline which are ranked with high scores. This indicates that the proposed method is reasonably robust in realistic operation conditions, i.e., does not produce an unreasonable amount of false positives.

\section{Summary and Conclusions}
\label{sec:conclusions}

We propose the application of relational learning on knowledge graphs to security monitoring and intrusion detection. The collective learning properties of graph embedding methods allow the resulting models to generalize beyond individual observations, benefiting from the context provided by a rich set of entity and relationship types. This in turn translates into an efficient use of training data, inherent robustness to false alarms in the presence of previously unobserved events, and potentially shorter baselining periods. 

The proposed method is experimentally evaluated in an industrial automation system prototype across a wide range of scenarios, and it is shown to leverage context information to produce a meaningful range of severity scores. This is particularly useful in the IDS setting, as observations are often not easily categorized as completely benign or malicious a priori, rather evoking varying degrees of interest or suspicion.

While these results are produced on a reduced-scale prototype, evaluation on a real-size multi-station manufacturing environment is underway. In addition, ongoing work aims to systematically assess the method's robustness to changing data modelling choices, and initial empirical results using various intuitive alternatives (e.g., representing network connections as either nodes or as edges connecting IP addresses) are encouraging, showing limited overall variations in detection performance, as improvements in some scenarios tend to compensate a slight degradation in others. 

Finally, future work includes creating synergistic combinations of the proposed graph-based method with other conventional solutions. Security analysts stand to benefit from a well-calibrated range of scores attached to IDS events, helping them focus their efforts on the most promising leads.

\section*{Acknowledgment}
The authors would like to thank their colleagues at the Semantics and Reasoning group and the Siemens AI Lab for the support and the many useful discussions in the course of this project. They would also like to thank Lukas Vogtmann for his assistance in building the evaluation prototype.

\printbibliography
\addcontentsline{toc}{section}{References}

\end{document}